\documentclass[amsmath,amssymb,11pt,reprint, preprintnumbers, notitlepage,aps,prl,twocolumn]{revtex4-1}
\pdfoutput=1 
\usepackage[utf8]{inputenc}
\usepackage[english]{babel}
\usepackage{amsmath}
\usepackage{graphicx}
\usepackage{dcolumn}
\usepackage{pbox}
\usepackage{amssymb}
\usepackage{epsfig}
\usepackage[dvipsnames]{xcolor}
\usepackage[normalem]{ulem}
\usepackage{slashed}
\usepackage{amssymb}
\usepackage{ mathrsfs }
\usepackage{color}
\usepackage[font=small]{caption}
\usepackage[font=small]{subcaption}
\usepackage{url}
\usepackage[makeroom]{cancel}
\usepackage{tikz}
\usetikzlibrary{arrows}
\usepackage{enumitem} 
\usepackage{soul}

\newcommand\tikzmark[1]{\tikz[remember picture,overlay]\coordinate (#1);}
\definecolor{MyDarkBlue}{rgb}{0.1, 0.1, 0.8} 
\definecolor{SBlue}{rgb}{0.2, 0.4, 0.7} 
\definecolor{MyLightBlue}{rgb}{0.22,0.51,0.9}
\definecolor{MyGreen}{rgb}{0.0, 0.5, 0.0}
\definecolor{BrickRed}{rgb}{0.8, 0.25, 0.33}
\RequirePackage{hyperref}
\hypersetup{colorlinks, citecolor=SBlue,linkcolor=MyDarkBlue, urlcolor=PineGreen}

\makeatletter
\renewcommand\@makecaption[2]{%
  \par
  \vskip\abovecaptionskip
  \begingroup
  
   \small\rmfamily
    \begingroup
     \samepage
     \flushing
     \let\footnote\@footnotemark@gobble
     \@make@capt@title{#1}{#2}\par
    \endgroup
  \endgroup
  \vskip\belowcaptionskip
}
\makeatother

\DeclareUnicodeCharacter{2212}{-}
\begin{document}
\title{\vspace{1cm}\Large 
Singling out SO(10) GUT models using recent PTA results
}

\author{\bf Stefan Antusch}
\email[E-mail:]{stefan.antusch@unibas.ch}
\author{\bf Kevin Hinze}
\email[E-mail:]{kevin.hinze@unibas.ch}
\author{\bf Shaikh Saad}
\email[E-mail:]{shaikh.saad@unibas.ch}
\author{\bf Jonathan Steiner}
\email[E-mail:]{jonathan.steiner@stud.unibas.ch}

\affiliation{Department of Physics, University of Basel, Klingelbergstrasse\ 82, CH-4056 Basel, Switzerland}

\begin{abstract}
In this work, we construct promising model building routes towards SO(10) GUT inflation and examine their ability to explain the recent PTA results hinting at a stochastic gravitational wave (GW) background at nanohertz frequencies. We consider a supersymmetric framework within which the so-called doublet-triplet splitting problem is solved without introducing fine-tuning. Additionally, realistic fermion masses and mixings, gauge coupling unification, and cosmic inflation are incorporated by utilizing superfields with representations no higher than the adjoint representation. Among the three possible scenarios, two of these cases require a single adjoint Higgs field, and do not lead to cosmic strings. In contrast, the third scenario featuring two adjoints, can lead to a network of metastable cosmic strings that generates a GW background contribution compatible with the recent PTA findings and testable by various ongoing and upcoming GW observatories.  
\end{abstract}

\maketitle
\textbf{Introduction:--}
Global collaboration among pulsar timing arrays (PTAs)  (NANOGrav~\cite{NANOGrav:2020bcs}, PPTA~\cite{Goncharov:2021oub}, EPTA~\cite{Chen:2021rqp}, and IPTA~\cite{Antoniadis:2022pcn}) previously revealed evidence of common-spectrum noise at nanohertz frequencies. Recent analysis, including CPTA~\cite{Xu:2023wog}, EPTA~\cite{Antoniadis:2023ott}, NANOGrav~\cite{NANOGrav:2023gor}, and PPTA~\cite{Reardon:2023gzh}, identified spatial correlations (Hellings-Downs effect~\cite{Hellings:1983fr}), providing strong support for a stochastic gravitational-wave background (SGWB). Although the mergers of supermassive black hole binaries (SMBHBs) are natural astrophysical sources of the SGWB at nanohertz frequencies, the new data somewhat disfavors SMBHBs in explaining the observed PTA SGWB signal~\cite{NANOGrav:2023gor}. Therefore, the SGWB likely points toward new physics beyond the Standard Model (SM). One of the explanations that fits well with the data is a metastable cosmic string network (CSN)~\cite{NANOGrav:2023hvm}. Since such cosmic strings (CSs) can arise from the multi-step spontaneous breaking of the symmetry group of a Grand Unified Theory (GUT) after cosmic inflation, this raises the question of what can be learned about GUTs from this finding.

GUTs~\cite{Pati:1973rp,Pati:1974yy, Georgi:1974sy, Georgi:1974yf, Georgi:1974my, Fritzsch:1974nn}, combined with SUSY, offer an appealing framework for a more fundamental theory beyond the SM of elementary particles. GUTs unify the three fundamental forces of the SM, while SUSY provides a natural solution to the gauge hierarchy problem and a potential weakly interacting dark matter candidate when R-parity or matter-parity ensures its stability. SO(10)-based GUTs are particularly interesting as they unify all SM fermions of each family into a single irreducible 16-dimensional representation. This 16-dimensional representation also includes a SM singlet right-handed neutrino, which, through the type-I seesaw mechanism~\cite{Minkowski:1977sc,Yanagida:1979as,Glashow:1979nm,Gell-Mann:1979vob,Mohapatra:1979ia}, generates tiny masses for the SM neutrinos.

Promising GUT models must satisfy proton decay bounds and achieve successful gauge coupling unification. In SUSY GUT models, the $d=5$ proton decay operators are induced by color-triplet exchange, necessitating the superheavy nature of color-triplet states compared to their doublet partners, known as the doublet-triplet splitting (DTS) problem~\cite{Randall:1995sh,Yamashita:2011an}. A desirable GUT model should solve the DTS problem without fine-tuning parameters. Since GUTs generate the Yukawa matrices out of joint GUT operators, leading to
constraints on the flavor structure, a further challenge
consists in realizing viable fermion masses and mixings.

Cosmic inflation~\cite{Guth:1980zm, Albrecht:1982wi, Linde:1981mu, Linde:1983gd} that solves the horizon and flatness problems of the standard Big Bang cosmology, and explains the origin of structure formation of the observable Universe, could have a deep connection to SUSY GUT models. In addition to the similarity of the scales of inflation and gauge coupling unification, inflation is also crucial to dilute away unwanted topological defects~\cite{Kibble:1976sj,Linde:1981mu} like monopoles which generically form at some stage of GUT symmetry breaking. Furthermore, supersymmetric theories typically possess many flat directions, providing an attractive framework for realizing inflation. While monopoles have to be diluted by inflation, other topological defects, like (metastable) CSs~\cite{Hindmarsh:1994re} that form after inflation can leave an observable signature in the SGWB.

In this work, we explore supersymmetric SO(10) GUTs that naturally solve the DTS problem, generate realistic fermion masses, and achieve successful gauge coupling unification and inflation. We focus on lower-dimensional field representations and investigate scenarios with Higgs fields no higher than the adjoint representation. Three promising routes for SO(10) GUT model building are identified: two cases use a single adjoint Higgs field, while the third scenario requires two copies. In the latter case, the intermediate symmetry contains two Abelian factors crucial for CSN formation. For the first time, we construct a realistic SUSY SO(10) GUT scenario (particularly the third scenario), satisfying the mentioned criteria and leads to metastable CSs capable of explaining the recent PTA results for a stochastic GW background at nanohertz frequencies.

\textbf{SO(10) model building:--}
Two major guiding principles in building realistic models in our framework are the natural DTS~\cite{Dimopoulos:1981xm,Srednicki:1982aj} (see also~\cite{Babu:1993we,Babu:1994kb,Berezhiani:1996bv,Barr:1997hq,Chacko:1998jz,Babu:1998wi,Babu:2002fsa,Kyae:2005vg,Babu:2010ej,Wan:2022glq}) and employing smaller dimensional representations. In achieving this, we  utilize $45_H$ and $16_H+\overline{16}_H$ Higgs representations to break the GUT symmetry down to the SM, which is subsequently broken by $10_H$ (and possibly by $16_H+\overline{16}_H$). The fundamental representation contains weak-doublet and color-triplet states,
\begin{align*}
10_H&=(2_H+3_H)+(\overline 2_H+\overline 3_H)
\nonumber\\
&=(1,2,1/2)+(3,1,-1/3)+c.c..     
\end{align*}
The VEV of the adjoint, $\langle 45_H\rangle\propto i\tau_2\otimes \mathrm{diag}(a_1,a_2,a_3,a_4,a_5)$ that breaks the GUT symmetry is expected to provide superheavy masses to both these components. With this setup, one can construct three classes of models: 
\begin{enumerate}[label=(\roman*)]
\item a single adjoint Higgs with $\langle 45_H\rangle\propto B-L$   generator,

\item a single adjoint Higgs with $\langle 45_H\rangle\propto I_{3R}$ generator,

\item two  adjoint Higgses, one with $\langle 45_H\rangle\propto B-L$   generator and another with $\langle 45_H^\prime\rangle\propto I_{3R}$ generator.
\end{enumerate}

For each model, the superpotential takes the form, 
\begin{align*}
W=&    W_\mathrm{GUT-breaking}+ \underbrace{ W_\mathrm{Inflation}+W_\mathrm{Mixed} }_{W_\mathrm{Intermedite-breaking}}
\\&
+W_\mathrm{DTS}+W_\mathrm{Yukawa},
\end{align*}
where terms in $W_\mathrm{GUT-breaking}$ and $W_\mathrm{Intermedite-breaking}$ lead to a consistent symmetry breaking of the GUT symmetry down to the SM gauge group.  Terms in $W_\mathrm{DTS}$ realize DTS without fine-tuning, and the $W_\mathrm{Inflation}$ part of the superpotential leads to an inflationary period.

$\bullet$\,\textbf{$B-L$-case:} The symmetry breaking chain in this scenario is given by
\begin{align*}
SO(10)
&\xrightarrow[45_H]{M_\mathrm{GUT}} 
SU(3)_{C}\times SU(2)_{L} \times SU(2)_{R}\times U(1)_{B-L} 
\\
&\xrightarrow[16_H+\overline{16}_H]{M_I} 
SU(3)_{C}\times SU(2)_{L} \times U(1)_{Y} \;.
\end{align*}
The GUT scale symmetry breaking is achieved via
\begin{align}
W_\mathrm{GUT-breaking}&\supset\frac{m_{45}}{2} \text{Tr}[45_H^2]+\frac{\lambda}{4\Lambda} \text{Tr}[45_H^4], \label{eq:200}   
\end{align}
with the VEV $\langle 45_H\rangle\propto i\tau_2\otimes \mathrm{diag}(a,a,a,0,0)$.

Note that breaking the GUT symmetry gives rise to superheavy monopoles that must be inflated away. Therefore inflation must take place after the formation of the monopoles. A straightforward option is to utilize hybrid~\cite{Linde:1993cn,linde1991axions,Dvali:1994ms} inflation (an alternative option is tribrid inflation \cite{Antusch:2004hd,Antusch:2010va}) at the last intermediate symmetry breaking stage, which we achieve via employing   $16_H+\overline{16}_H$ that acquire VEVs~\footnote{As a result, the appearance of automatic R-parity from within the SO(10) group is no longer possible.  However a discrete symmetry, such as a $Z_2$ symmetry (matter parity), can readily be imposed.} in the right-handed neutrino direction. Then the relevant superpotential term contributing to inflation takes the following form, 
\begin{align}
W_\mathrm{Inflation} \supset \kappa S(\overline{16}_H16_H-m_{16}^2),    
\end{align}
which fixes the magnitude of the VEVs $\langle 16_H\overline{16}_H\rangle = m^2_{16}$. 
Here, $S$ is a GUT singlet superfield, the scalar component of which plays the role of the inflaton.

Since $45_H$ and  $16_H+\overline{16}_H$ have component fields that share the same quantum numbers, 
\begin{align*}
45_H, 16_H, \overline{16}_H\supset (1,1,1)+ (3,2,1/6)+ (\overline 3,1,-2/3) +c.c.,    
\end{align*}
to avoid additional would-be Goldstone bosons, which would ruin gauge coupling unification, these fields must have non-trivial mixing terms. The simplest possible interaction term, $\overline{16}_H 45_H 16_H$,  is not welcome since it would destabilize the VEV of $45_H$ from the desired ``Dimopoulos-Wilczek form''.

To circumvent this issue, we introduce a second copy of spinorial representations, $16_H^\prime+\overline{16}_H^\prime$, which do not acquire a VEV in the  right-handed neutrino direction. Then a consistent symmetry breaking without additional would-be Goldstone bosons can be achieved via the addition of the following terms in the superpotential:
\begin{align}
&W_\mathrm{Mixed}\supset
\\&
\overline{16}_H(\lambda_1 45_H+\lambda^\prime_1 1_H)16_H^\prime +\overline{16}_H^\prime (\lambda_2 45_H+\lambda^\prime_2 1^\prime_H)16_H. \nonumber
\end{align}
Here, we introduced the ``sliding singlets'' $1^{(\prime)}_H$, which  are assumed to have no other terms in the superpotential that could fix their VEVs. They are needed to allow for vanishing $F$-terms corresponding to $16_H^\prime, \overline{16}_H^\prime$.

Concerning DTS, remarkably, the specific VEV structure of the $45_H$ provides masses to only the color-triplets,  while the weak-doublets remain massless, schematically
\begin{align}
&10_{1H} \langle 45_H\rangle 10_{2H}= \nonumber \label{eq:100}\\& 
\cancelto{0}{\overline 2_{1H} 2_{2H}} 
+\cancelto{0}{\overline 2_{2H} 2_{1H}}
+\overline 3_{1H} 3_{2H} 
+\overline 3_{2H} 3_{1H}.
\end{align}
However, if only the above term is added to the superpotential, then the low energy spectrum would contain four light doublets instead of the usual two doublets of the MSSM. This would spoil the successful gauge coupling unification of the MSSM.  To avoid extra light states, we allow a direct mass term for $10_{2H}$, i.e.,
\begin{align}
10_{2H}10_{2H}= \label{eq:101}
\overline 2_{2H} 2_{2H}
+\overline 3_{2H} 3_{2H}. 
\end{align}
Then, the terms in the superpotential relevant for providing the masses of the doublets and triplets and naturally realizing their splittings are
\begin{align}
&W_\mathrm{DTS}\supset \label{eq:102}
\gamma 10_{1H} 45_H 10_{2H} +m_{10} 10_{2H} 10_{2H}.
\end{align}

A crucial remark is in order. Assuming that only $10_{1H}$ couples to the fermions, the term in Eq.~\eqref{eq:100} by itself does not induce proton decay. Once the term in Eq.~\eqref{eq:101} is also introduced, together they allow the proton to decay via color-triplet Higgses, since now an effective mass term linking $\overline 3_{1H}$ and $3_{1H}$ can be written down after integrating out $\overline 3_{2H}$ and $3_{2H}$. This can be understood schematically as follows: 
\[
\overline 3_{1H} \langle 45_H\rangle {\tikzmark{z1}3_{2H}}
\hspace{20pt}
{\tikzmark{z2} \overline 3_{2H}} m_{10} {\tikzmark{z3}3_{2H}}
\hspace{20pt}
{\tikzmark{z4}\overline 3_{2H}} \langle 45_H\rangle 3_{1H} \;. 
\]

\begin{tikzpicture}[remember picture,overlay]
  \draw[latex-latex]
  ([shift={(2pt,-2pt)}]z1)
  -- ([shift={(2pt,-12pt)}]z1)
  -- node[midway, below] {$\scriptstyle$} ([shift={(2pt,-12pt)}]z2)
  -- ([shift={(2pt,-2pt)}]z2);

  \draw[latex-latex]
  ([shift={(2pt,-2pt)}]z3)
  -- ([shift={(2pt,-12pt)}]z3)
  -- node[midway, below] {$\scriptstyle$} ([shift={(2pt,-12pt)}]z4)
  -- ([shift={(2pt,-2pt)}]z4);
\end{tikzpicture}\\
\noindent With a sufficiently large effective triplet mass $\sim M^2_\mathrm{GUT}/m_{10}$, the $d=5$ proton decay is suppressed.

$\bullet$\,\textbf{$I_{3R}$-case:} The symmetry breaking chain in this scenario is given by
\begin{align*}
SO(10)
&\xrightarrow[45_H]{M_\mathrm{GUT}} 
SU(4)_{C}\times SU(2)_{L} \times U(1)_{R} 
\\
&\xrightarrow[16_H+\overline{16}_H]{M_I} 
SU(3)_{C}\times SU(2)_{L} \times U(1)_{Y}  \;,
\end{align*}
which is obtained by $\langle 45_H\rangle\propto i\tau_2\otimes \mathrm{diag}(0,0,0,b,b)$. Although the $W_\mathrm{GUT-breaking}$ and $W_\mathrm{Intermedite-breaking}$  parts of the superpotential are identical to the $B-L$ case, $W_\mathrm{DTS}$ takes a different form, which we discuss in the following. 

Due to $\langle 45_H\rangle\propto I_{3R}$, we now have the opposite situation compared to the previous case, namely
\begin{align}
&10_{1H} \langle 45_H\rangle 10_{2H}= 
\nonumber\\
&\overline 2_{1H} 2_{2H} 
+\overline 2_{2H} 2_{1H}
+\cancelto{0}{\overline 3_{1H} 3_{2H}} 
+\cancelto{0}{\overline 3_{2H} 3_{1H}}.
\end{align}
Therefore, a different strategy must be implemented to obtain light doublets and superheavy color-triplets.  By noting that
 $16_H^\prime\supset\overline 2_H^\prime$ is a $SU(2)_R$ singlet, and, on the contrary, $16_H^\prime\supset \overline 3_H^\prime$ resides in a $SU(2)_R$ doublet, one obtains a mass only for the color-triplet, and not for the weak doublet, i.e.,
\begin{align}
\overline{16}_{H}^\prime\langle 45_H\rangle 16_{H}^\prime= 
\cancelto{0}{\overline 2_{H}^\prime 2_{H}^\prime} 
+\overline 3_{H}^\prime 3_{H}^\prime \;. 
\end{align}
If only the above term is included in the superpotential, then a pair of triplets will remain massless in addition to one pair of doublets. To provide large masses to all the color-triplets, we add two more terms 
\begin{align}
&W_\mathrm{DTS}\supset \\&
\lambda_3\overline{16}_H^\prime 45_H 16^\prime_H
 +\lambda_4 10_H 16_H 16_H+\lambda_5  10_H \overline{16}_H  \overline{16}_H \;.\nonumber
\end{align}

As for the $d=5$ proton decay, 
assuming the SM fermion masses are coming from their coupling to the $10_H$ (i.e. neglecting all contributions from the $16_H$), the effective triplet mass $m_T$ is approximately given by 
\begin{align}
    m_T=-\frac{\lambda_3\lambda_4\lambda_5\langle 16_H\rangle \langle \overline{16}_H\rangle}{2\lambda_1\lambda_2 \langle 45_H\rangle}.
\end{align}
Choosing somewhat small $\lambda_1,\lambda_2$ allows having $m_T\gtrsim 10^{19}$~GeV, which is required by proton decay constraints.

$\bullet$\,\textbf{$B-L \;\&\; I_{3R}$-case:} Depending on the values of the VEVs of the two adjoints, various symmetry breaking chains may arise in this scenario, examples of which are  (a) $\langle 45_H\rangle > \langle 45_H^\prime\rangle > \langle 16_H\rangle, \langle \overline{16}_H\rangle$:
\begin{align*}
SO(10)
&\xrightarrow[45_H]{M_\mathrm{GUT}} 
SU(3)_{C}\times SU(2)_{L} \times SU(2)_{R}\times U(1)_{B-L} 
\\
&\xrightarrow[45_H^\prime]{M_I} 
SU(3)_{C}\times SU(2)_{L} \times U(1)_{R}  \times U(1)_{B-L} 
\\
&\xrightarrow[16_H+\overline{16}_H]{M_{II}} 
SU(3)_{C}\times SU(2)_{L} \times U(1)_{Y}  \;,
\end{align*}
(b) $\langle 45_H^\prime\rangle > \langle 45_H\rangle > \langle 16_H\rangle, \langle \overline{16}_H\rangle$:
\begin{align*}
SO(10)
&\xrightarrow[45_H^\prime]{M_\mathrm{GUT}} 
SU(4)_{C}\times SU(2)_{L} \times U(1)_{R} 
\\
&\xrightarrow[45_H]{M_I} 
SU(3)_{C}\times SU(2)_{L} \times U(1)_{R}  \times U(1)_{B-L} 
\\
&\xrightarrow[16_H+\overline{16}_H]{M_{II}} 
SU(3)_{C}\times SU(2)_{L} \times U(1)_{Y}  \;,
\end{align*}
(c) $\langle 45_H\rangle = \langle 45_H^\prime\rangle > \langle 16_H\rangle, \langle \overline{16}_H\rangle$:
\begin{align*}
SO(10)
&\xrightarrow[45_H+45_H^\prime]{M_\mathrm{GUT}} 
SU(3)_{C}\times SU(2)_{L} \times U(1)_{R}  \times U(1)_{B-L} 
\\
&\xrightarrow[16_H+\overline{16}_H]{M_{I}} 
SU(3)_{C}\times SU(2)_{L} \times U(1)_{Y}  \;.
\end{align*}

In this scenario, for each of the adjoints, the GUT symmetry breaking superpotential consists of the terms given in Eq.~\eqref{eq:200}. Since $\langle 45_H\rangle$ and $\langle 45_H^\prime\rangle$ break SO(10) to the left-right symmetry and quark-lepton symmetry, respectively, the first and the second break the generators in $(3,2,+1/6)+(3,2,-5/6)+(3,1,2/3)+c.c$ and $(3,2,+1/6)+(3,2,-5/6)+(1,1,+1)+c.c$, respectively. Consequently, there would be additional  massless states.  To avoid such massless states, we add the following mixing term in the superpotential, 
\begin{align}
W_\mathrm{GUT-breaking}&\supset \frac{\eta}{\Lambda} Tr[45_H.45_H.45_H^\prime .45_H^\prime].
\end{align}

As before, one requires non-trivial interactions between the spinorial representations and the adjoints to give masses to the would-be Goldstones. For the two adjoints, we now introduce two sets of additional spinorial representations, $16_H^\prime+\overline{16}_H^\prime$ and $16_H^{\prime\prime}+\overline{16}_H^{\prime\prime}$, and add the following terms, such that the VEVs of the adjoints are not destabilized: 
\begin{align}
&W_\mathrm{Mixed}\supset 
\\&
\overline{16}_H(\lambda_1 45_H+\lambda_1^\prime 1_H)16_H^\prime +\overline{16}_H^\prime (\lambda_2 45_H+\lambda_2^\prime 1_H^\prime)16_H
\nonumber\\
+&
\overline{16}_H(\lambda_3 45_H^\prime+\lambda_3^\prime 1_H^{\prime\prime})16_H^{\prime\prime} +\overline{16}_H^{\prime\prime} (\lambda_4 45_H^\prime+\lambda_4^\prime 1_H^{\prime\prime\prime})16_H .   \nonumber 
\end{align}

For the DTS, we include the term $10_{1H}45_H10_{2H}$. However, here we can construct an example model which does not lead to proton decay at leading order via $d=5$ operators. To this end, we forbid the direct mass term $10_{2H}10_{2H}$. Instead, we include a higher dimensional operator, $10_{2H}. 45^{\prime 2}.10_{2H}$, such that an effective triplet mass for $3_{1H}$ and $\overline 3_{1H}$ cannot be written down, since,
\begin{align}
&10_{2H}45_H^{\prime 2}10_{2H}= 
\nonumber \\&
\overline 2_{2H} 2_{2H} 
+\overline 2_{2H} 2_{2H}
+\cancelto{0}{\overline 3_{2H} 3_{2H}} 
+\cancelto{0}{\overline 3_{2H} 3_{2H}}.
\end{align}
With the inclusion of the above two terms, still one pair of color-triplets and an additional pair of weak doublets remain massless. We cure this by adding a term of the form $\overline{16}^{\prime\prime}_H 16^\prime_H$ to the superpotential,
\begin{align}
&W_\mathrm{DTS} \supset 
\\&
\gamma_1 10_{1H} 45_H 10_{2H} +\frac{\gamma_2}{\Lambda} 10_{2H}45_H^{\prime 2}10_{2H}
+\omega_{16} \overline{16}^{\prime\prime}_H 16^\prime_H ,\nonumber 
\end{align}
that leads to a single pair of light doublets, as desired.

It is important to note that all the scenarios discussed above can successfully reproduce correct charged fermion masses and mixings by incorporating suitable higher-dimensional operators. The light neutrinos acquire masses through the standard type-I seesaw mechanism. The Majorana masses for the right-handed neutrinos are generated by the following higher-dimensional operator:
\begin{align}\label{eq:RHneutrinomasses}
W_\mathrm{Yukawa}\supset Y_R 16_i 16_j \frac{\overline{16}_H\overline{16}_H}{\Lambda} \sim Y_R \frac{v^2_R}{\Lambda} \nu^c\nu^c \:.
\end{align}

\textbf{Gravitational wave signals:--}
In some of the models we consider, breaking e.g.\ a simple group into a subgroup that contains a $U(1)$ factor leads to monopole creation. To prevent overclosing the universe, inflation must get rid of the monopoles. At some later stage, once the left-over Abelian symmetry is broken, strings appear (we assume the ideal Nambu-Goto string approximation, where the dominant radiation emission of CSs is into GWs~\cite{Vachaspati:1984gt}).  If these two scales are very close, Schwinger nucleation of monopole-antimonopole pairs~\cite{Langacker:1980kd,Lazarides:1981fv,Vilenkin:1982hm} on the string cuts it into pieces and makes it decay. How quickly these metastable strings decay depends on a parameter $\kappa_m$~\cite{Leblond:2009fq},
\begin{align}
\kappa_m= \frac{m^2}{\mu}\sim \frac{8\pi}{g^2} \left( \frac{v_m}{v_R} \right)^2,    
\end{align}
where $m$ is the mass of the monopole and $v_m$ ($v_R$) is the monopole (string) creation scale. The network behaves like a stable-string network for $\kappa_m^{1/2}\gg 10$.

\begin{figure}[t!]
\begin{center}
\includegraphics[width=0.48\textwidth]{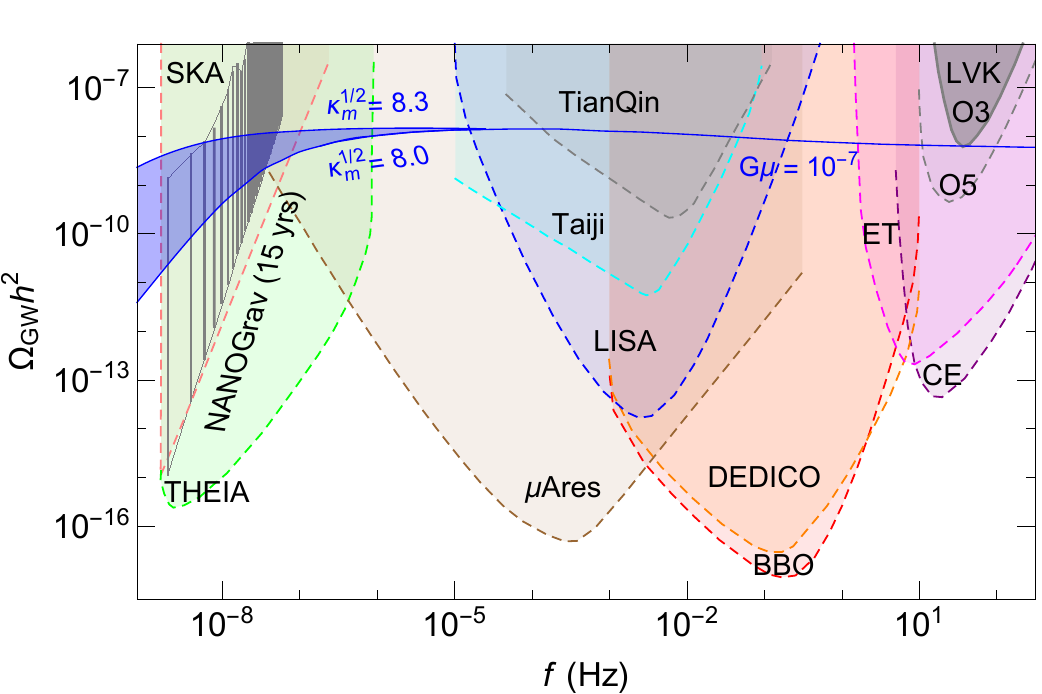}
\caption{Examples of GW signals (which we estimate following~\cite{Buchmuller:2021mbb}) from metastable cosmic strings explaining the recent PTA result~\cite{NANOGrav:2023hvm} while being at the edge of the sensitivity reach of LVK~\cite{LIGOScientific:2021nrg,KAGRA:2021kbb}. The plot also shows examples of sensitivities of possible future observatories (SKA~\cite{Janssen:2014dka}, THEIA~\cite{Theia:2017xtk},
$\mu$Ares~\cite{Sesana:2019vho}, LISA~\cite{Audley:2017drz}, Taiji~\cite{Guo:2018npi}, TianQin~\cite{Luo:2015ght}, BBO~\cite{Corbin:2005ny}, DECIGO~\cite{Seto:2001qf}, ET~\cite{Sathyaprakash:2012jk} and CE~\cite{Evans:2016mbw}) which can test the signal at various frequencies. Moreover, the recent NANOGrav 15 yrs result is also shown with grey lines. $f$ is the GW frequency observed today.
}\label{fig:fig-01}
\end{center}
\end{figure}

Metastable CSNs provide an intriguing explanation for the newly released PTA data~\cite{NANOGrav:2023hvm}. The data indicates string tension ($\mu$) values in the range $G\mu\sim 10^{-8}-10^{-5}$ for $\kappa_m^{1/2}\sim 7.7-8.3$ (with a strong correlation, cf.\ Fig.\ 10 of~\cite{NANOGrav:2023hvm}), consistent with CMB bounds. Notably, the $68\%$ credible region in the $G\mu-\kappa_m^{1/2}$ parameter plane overlaps with the third advanced LIGO–Virgo–KAGRA (LVK) bound, while major parts of the $95\%$ credible region are compatible, preferring  $G\mu\lesssim 10^{-7}$ and $\kappa_m^{1/2}\sim 8$~\cite{NANOGrav:2023hvm}, as shown in Fig.\ref{fig:fig-01}. However, it should be remarked that the computation of the GW spectrum from metastable CSs carries significant uncertainty~\cite{Auclair:2019wcv}. Furthermore, various possible effects are not included  in the above shown GW spectrum, for instance, an extended matter domination phase after inflation~\cite{Cui:2018rwi,Auclair:2019wcv,Blasi:2020wpy} or the change of degrees of freedom below the SUSY breaking scale~\cite{Cui:2018rwi}. Nevertheless, observing a higher frequency SGWB signal in the next LIGO–Virgo–KAGRA rounds would be a fascinating confirmation of the scenario.

Interestingly, $G\mu\sim 10^{-7}$ corresponds roughly to $v_R\sim 10^{15}$ GeV, which is fully consistent with the type-I seesaw contribution to neutrino masses and corresponds to the right scale for inflation. On the other hand, stable CSs are disfavored by the recent PTA data\footnote{Stable cosmic strings, however, were consistent with the previous PTA data. For works on GWs, in light of NANOGrav12.5 data, arising from cosmic strings within GUTs, c.f.,~\cite{Buchmuller:2019gfy,King:2020hyd,King:2021gmj,Lazarides:2022jgr,Fu:2022lrn,Saad:2022mzu,Lazarides:2023iim,Maji:2023fba,Madge:2023cak}.}.

The first (and second) model studied, the $B-L$- (and $I_{3R}$-) case, leads to embedded strings, which are generally unstable~\cite{James:1992zp, James:1992wb, Goodband:1995he}. Interestingly, all three models in the $B-L \;\&\; I_{3R}$-case have the potential to produce metastable strings for nearly degenerate monopole and string formation scales: $M_I\sim M_{II}$ for cases (a) and (b), and $M_\mathrm{GUT}\sim M_I$ for case (c). However, in case (c), a lower GUT scale $\sim 10^{15}$ GeV would have to be arranged that requires suppression of $d=6$ proton decay utilizing the freedom in the Yukawa sector, which makes this case somewhat less appealing.   
We like to point out that the class of promising SO(10) models we considered in this work may or may not lead to the formation of CSs, contrary to the class of models considered in~\cite{Jeannerot:2003qv}, where the appearance of CSs is unavoidable.

Before concluding, we discuss the gauge coupling unification for an example scenario that leads to metastable CSs (specifically, we choose case (a) within $B-L\;\&\;I_{3R}$). To achieve metastable strings, the monopole and string formation scales must nearly coincide. Therefore, we effectively have three scales: the GUT scale, the monopole/string formation scale, and the SUSY breaking scale (fixed at 3 TeV). To simplify the analysis, we assume that the fields breaking a symmetry are degenerate with the corresponding scale, while the remaining states have GUT scale masses. This minimal number of free parameters allows us to find a wide range for the monopole/string formation scale, approximately $M_I\sim M_{II}\sim [10^9-10^{17}]$ GeV (with $10^{16}$ GeV $\leq M_\mathrm{GUT} \leq 10^{18}$ GeV and $M_\mathrm{GUT}>M_I$), while still being consistent with gauge coupling unification. Our analysis considers a $1\%$ uncertainty on the measured values of the gauge couplings to account for GUT threshold uncertainties.

A comprehensive analysis encompassing gauge coupling unification, fermion masses and mixings, proton decay, GW signal, and the mass spectrum of the component fields from the superpotential terms will be presented in a forthcoming publication.

\textbf{Conclusions:--}
We explored promising model-building routes for SO(10) GUT inflation in light of the recent PTA results suggesting the presence of a SGWB at nanohertz frequencies. Our investigation focused on a supersymmetric SO(10) framework with small dimensional representations, effectively solving the doublet-triplet splitting problem without fine-tuning. This approach enables realistic fermion masses, gauge coupling unification, and simple options for embedding cosmic inflation. Among the three model classes studied, one involves two adjoint fields capable of generating a network of metastable cosmic strings. This network generates a SGWB background contribution that can explain the recent PTA data, and will be tested by various upcoming GW observatories.

\vspace{20pt}
\text{\bf Note added:} As we were completing this work, several papers appeared that also discussed the
impact of the new PTA results on new physics scenarios~\cite{Zu:2023olm,Vagnozzi:2023lwo,Han:2023olf,Lambiase:2023pxd,Ellis:2023dgf,Guo:2023hyp,Megias:2023kiy,Fujikura:2023lkn,Yang:2023aak,Franciolini:2023wjm,Shen:2023pan,Kitajima:2023cek,Ellis:2023tsl,Franciolini:2023pbf,Ghoshal:2023fhh,Bai:2023cqj,Addazi:2023jvg,Athron:2023mer,Kitajima:2023vre,Lazarides:2023ksx,Broadhurst:2023tus,Cai:2023dls,Yang:2023qlf,Blasi:2023sej,Inomata:2023zup,Depta:2023qst,Gouttenoire:2023ftk,Borah:2023sbc,Murai:2023gkv,Datta:2023vbs,Barman:2023fad,Bi:2023tib,Lu:2023mcz,Xiao:2023dbb,Li:2023bxy,Anchordoqui:2023tln,Niu:2023bsr,Ebadi:2023xhq,Gouttenoire:2023nzr,Abe:2023yrw,Ghosh:2023aum,Unal:2023srk,Bian:2023dnv,Figueroa:2023zhu,Du:2023qvj,Servant:2023mwt,Li:2023tdx,DiBari:2023upq,Geller:2023shn,Liu:2023ymk}.

\bibliographystyle{style}
\bibliography{reference}

\providecommand{\href}[2]{#2}\begingroup\raggedright\begin{thebibliography}{100}

\bibitem{NANOGrav:2020bcs}
{\bfseries NANOGrav} Collaboration, Z.~Arzoumanian {\em et~al.}, ``{The
  NANOGrav 12.5 yr Data Set: Search for an Isotropic Stochastic
  Gravitational-wave Background},''
  \href{http://dx.doi.org/10.3847/2041-8213/abd401}{{\em Astrophys. J. Lett.}
  {\bfseries 905} no.~2, (2020) L34},
  \href{http://arxiv.org/abs/2009.04496}{{\ttfamily arXiv:2009.04496
  [astro-ph.HE]}}.

\bibitem{Goncharov:2021oub}
B.~Goncharov {\em et~al.}, ``{On the Evidence for a Common-spectrum Process in
  the Search for the Nanohertz Gravitational-wave Background with the Parkes
  Pulsar Timing Array},''
  \href{http://dx.doi.org/10.3847/2041-8213/ac17f4}{{\em Astrophys. J. Lett.}
  {\bfseries 917} no.~2, (2021) L19},
  \href{http://arxiv.org/abs/2107.12112}{{\ttfamily arXiv:2107.12112
  [astro-ph.HE]}}.

\bibitem{Chen:2021rqp}
S.~Chen {\em et~al.}, ``{Common-red-signal analysis with 24-yr high-precision
  timing of the European Pulsar Timing Array: inferences in the stochastic
  gravitational-wave background search},''
  \href{http://dx.doi.org/10.1093/mnras/stab2833}{{\em Mon. Not. Roy. Astron.
  Soc.} {\bfseries 508} no.~4, (2021) 4970--4993},
  \href{http://arxiv.org/abs/2110.13184}{{\ttfamily arXiv:2110.13184
  [astro-ph.HE]}}.

\bibitem{Antoniadis:2022pcn}
J.~Antoniadis {\em et~al.}, ``{The International Pulsar Timing Array second
  data release: Search for an isotropic gravitational wave background},''
  \href{http://dx.doi.org/10.1093/mnras/stab3418}{{\em Mon. Not. Roy. Astron.
  Soc.} {\bfseries 510} no.~4, (2022) 4873--4887},
  \href{http://arxiv.org/abs/2201.03980}{{\ttfamily arXiv:2201.03980
  [astro-ph.HE]}}.

\bibitem{Xu:2023wog}
H.~Xu {\em et~al.}, ``{Searching for the Nano-Hertz Stochastic Gravitational
  Wave Background with the Chinese Pulsar Timing Array Data Release I},''
  \href{http://dx.doi.org/10.1088/1674-4527/acdfa5}{{\em Res. Astron.
  Astrophys.} {\bfseries 23} no.~7, (2023) 075024},
  \href{http://arxiv.org/abs/2306.16216}{{\ttfamily arXiv:2306.16216
  [astro-ph.HE]}}.

\bibitem{Antoniadis:2023ott}
J.~Antoniadis {\em et~al.}, ``{The second data release from the European Pulsar
  Timing Array III. Search for gravitational wave signals},''
  \href{http://arxiv.org/abs/2306.16214}{{\ttfamily arXiv:2306.16214
  [astro-ph.HE]}}.

\bibitem{NANOGrav:2023gor}
{\bfseries NANOGrav} Collaboration, G.~Agazie {\em et~al.}, ``{The NANOGrav 15
  yr Data Set: Evidence for a Gravitational-wave Background},''
  \href{http://dx.doi.org/10.3847/2041-8213/acdac6}{{\em Astrophys. J. Lett.}
  {\bfseries 951} no.~1, (2023) L8},
  \href{http://arxiv.org/abs/2306.16213}{{\ttfamily arXiv:2306.16213
  [astro-ph.HE]}}.

\bibitem{Reardon:2023gzh}
D.~J. Reardon {\em et~al.}, ``{Search for an Isotropic Gravitational-wave
  Background with the Parkes Pulsar Timing Array},''
  \href{http://dx.doi.org/10.3847/2041-8213/acdd02}{{\em Astrophys. J. Lett.}
  {\bfseries 951} no.~1, (2023) L6},
  \href{http://arxiv.org/abs/2306.16215}{{\ttfamily arXiv:2306.16215
  [astro-ph.HE]}}.

\bibitem{Hellings:1983fr}
R.~w. Hellings and G.~s. Downs, ``{Upper Limits on the Isotropic Gravitational
  Radition Background from Pulsar Timing Analysis},''
  \href{http://dx.doi.org/10.1086/183954}{{\em Astrophys. J. Lett.} {\bfseries
  265} (1983) L39--L42}.

\bibitem{NANOGrav:2023hvm}
{\bfseries NANOGrav} Collaboration, A.~Afzal {\em et~al.}, ``{The NANOGrav 15
  yr Data Set: Search for Signals from New Physics},''
  \href{http://dx.doi.org/10.3847/2041-8213/acdc91}{{\em Astrophys. J. Lett.}
  {\bfseries 951} no.~1, (2023) L11},
  \href{http://arxiv.org/abs/2306.16219}{{\ttfamily arXiv:2306.16219
  [astro-ph.HE]}}.

\bibitem{Pati:1973rp}
J.~C. Pati and A.~Salam, ``{Is Baryon Number Conserved?},''
  \href{http://dx.doi.org/10.1103/PhysRevLett.31.661}{{\em Phys. Rev. Lett.}
  {\bfseries 31} (1973) 661--664}.

\bibitem{Pati:1974yy}
J.~C. Pati and A.~Salam, ``{Lepton Number as the Fourth Color},''
  \href{http://dx.doi.org/10.1103/PhysRevD.10.275}{{\em Phys. Rev. D}
  {\bfseries 10} (1974) 275--289}. [Erratum: Phys.Rev.D 11, 703--703 (1975)].

\bibitem{Georgi:1974sy}
H.~Georgi and S.~L. Glashow, ``{Unity of All Elementary Particle Forces},''
\href{http://dx.doi.org/10.1103/PhysRevLett.32.438}{{\em Phys. Rev. Lett.}
  {\bfseries 32} (1974) 438--441}.

\bibitem{Georgi:1974yf}
H.~Georgi, H.~R. Quinn, and S.~Weinberg, ``{Hierarchy of Interactions in
  Unified Gauge Theories},''
  \href{http://dx.doi.org/10.1103/PhysRevLett.33.451}{{\em Phys. Rev. Lett.}
  {\bfseries 33} (1974) 451--454}.

\bibitem{Georgi:1974my}
H.~Georgi, ``{The State of the Art\textemdash{}Gauge Theories},''
  \href{http://dx.doi.org/10.1063/1.2947450}{{\em AIP Conf. Proc.} {\bfseries
  23} (1975) 575--582}.

\bibitem{Fritzsch:1974nn}
H.~Fritzsch and P.~Minkowski, ``{Unified Interactions of Leptons and
  Hadrons},'' \href{http://dx.doi.org/10.1016/0003-4916(75)90211-0}{{\em Annals
  Phys.} {\bfseries 93} (1975) 193--266}.

\bibitem{Minkowski:1977sc}
P.~Minkowski, ``{$\mu \to e\gamma$ at a Rate of One Out of $10^{9}$ Muon
  Decays?},''
\href{http://dx.doi.org/10.1016/0370-2693(77)90435-X}{{\em Phys. Lett.}
  {\bfseries 67B} (1977) 421--428}.

\bibitem{Yanagida:1979as}
T.~Yanagida, ``{Horizontal gauge symmetry and masses of neutrinos},''
{\em Conf. Proc.} {\bfseries C7902131} (1979) 95--99.

\bibitem{Glashow:1979nm}
S.~Glashow, ``{The Future of Elementary Particle Physics},''
  \href{http://dx.doi.org/10.1007/978-1-4684-7197-7\_15}{{\em NATO Sci. Ser. B}
  {\bfseries 61} (1980) 687}.

\bibitem{Gell-Mann:1979vob}
M.~Gell-Mann, P.~Ramond, and R.~Slansky, ``{Complex Spinors and Unified
  Theories},'' {\em Conf. Proc. C} {\bfseries 790927} (1979) 315--321,
  \href{http://arxiv.org/abs/1306.4669}{{\ttfamily arXiv:1306.4669 [hep-th]}}.

\bibitem{Mohapatra:1979ia}
R.~N. Mohapatra and G.~Senjanovic, ``{Neutrino Mass and Spontaneous Parity
  Nonconservation},'' \href{http://dx.doi.org/10.1103/PhysRevLett.44.912}{{\em
  Phys. Rev. Lett.} {\bfseries 44} (1980) 912}.

\bibitem{Randall:1995sh}
L.~Randall and C.~Csaki, ``{The Doublet - triplet splitting problem and Higgses
  as pseudoGoldstone bosons},'' in {\em {International Workshop on
  Supersymmetry and Unification of Fundamental Interactions (SUSY 95)}},
  pp.~99--109.
\newblock 3, 1995.
\newblock \href{http://arxiv.org/abs/hep-ph/9508208}{{\ttfamily
  arXiv:hep-ph/9508208}}.

\bibitem{Yamashita:2011an}
T.~Yamashita, ``{Doublet-Triplet Splitting in an SU(5) Grand Unification},''
  \href{http://dx.doi.org/10.1103/PhysRevD.84.115016}{{\em Phys. Rev. D}
  {\bfseries 84} (2011) 115016},
  \href{http://arxiv.org/abs/1106.3229}{{\ttfamily arXiv:1106.3229 [hep-ph]}}.

\bibitem{Guth:1980zm}
A.~H. Guth, ``{The Inflationary Universe: A Possible Solution to the Horizon
  and Flatness Problems},''
  \href{http://dx.doi.org/10.1103/PhysRevD.23.347}{{\em Phys. Rev. D}
  {\bfseries 23} (1981) 347--356}.

\bibitem{Albrecht:1982wi}
A.~Albrecht and P.~J. Steinhardt, ``{Cosmology for Grand Unified Theories with
  Radiatively Induced Symmetry Breaking},''
  \href{http://dx.doi.org/10.1103/PhysRevLett.48.1220}{{\em Phys. Rev. Lett.}
  {\bfseries 48} (1982) 1220--1223}.

\bibitem{Linde:1981mu}
A.~D. Linde, ``{A New Inflationary Universe Scenario: A Possible Solution of
  the Horizon, Flatness, Homogeneity, Isotropy and Primordial Monopole
  Problems},'' \href{http://dx.doi.org/10.1016/0370-2693(82)91219-9}{{\em Phys.
  Lett.} {\bfseries 108B} (1982) 389--393}.
[Adv. Ser. Astrophys. Cosmol.3,149(1987)].

\bibitem{Linde:1983gd}
A.~D. Linde, ``{Chaotic Inflation},''
  \href{http://dx.doi.org/10.1016/0370-2693(83)90837-7}{{\em Phys. Lett. B}
  {\bfseries 129} (1983) 177--181}.

\bibitem{Kibble:1976sj}
T.~W.~B. Kibble, ``{Topology of Cosmic Domains and Strings},''
  \href{http://dx.doi.org/10.1088/0305-4470/9/8/029}{{\em J. Phys. A}
  {\bfseries 9} (1976) 1387--1398}.

\bibitem{Hindmarsh:1994re}
M.~B. Hindmarsh and T.~W.~B. Kibble, ``{Cosmic strings},''
  \href{http://dx.doi.org/10.1088/0034-4885/58/5/001}{{\em Rept. Prog. Phys.}
  {\bfseries 58} (1995) 477--562},
  \href{http://arxiv.org/abs/hep-ph/9411342}{{\ttfamily arXiv:hep-ph/9411342}}.

\bibitem{Dimopoulos:1981xm}
S.~Dimopoulos and F.~Wilczek, ``{Incomplete Multiplets in Supersymmetric
  Unified Models},''.

\bibitem{Srednicki:1982aj}
M.~Srednicki, ``{Supersymmetric Grand Unified Theories and the Early
  Universe},'' \href{http://dx.doi.org/10.1016/0550-3213(82)90073-6}{{\em Nucl.
  Phys. B} {\bfseries 202} (1982) 327--335}.

\bibitem{Babu:1993we}
K.~S. Babu and S.~M. Barr, ``{Natural suppression of Higgsino mediated proton
  decay in supersymmetric SO(10)},''
  \href{http://dx.doi.org/10.1103/PhysRevD.48.5354}{{\em Phys. Rev. D}
  {\bfseries 48} (1993) 5354--5364},
  \href{http://arxiv.org/abs/hep-ph/9306242}{{\ttfamily arXiv:hep-ph/9306242}}.

\bibitem{Babu:1994kb}
K.~S. Babu and R.~N. Mohapatra, ``{Mass matrix textures from superstring
  inspired SO(10) models},''
  \href{http://dx.doi.org/10.1103/PhysRevLett.74.2418}{{\em Phys. Rev. Lett.}
  {\bfseries 74} (1995) 2418--2421},
  \href{http://arxiv.org/abs/hep-ph/9410326}{{\ttfamily arXiv:hep-ph/9410326}}.

\bibitem{Berezhiani:1996bv}
Z.~Berezhiani and Z.~Tavartkiladze, ``{More missing VEV mechanism in
  supersymmetric SO(10) model},''
  \href{http://dx.doi.org/10.1016/S0370-2693(97)00873-3}{{\em Phys. Lett. B}
  {\bfseries 409} (1997) 220--228},
  \href{http://arxiv.org/abs/hep-ph/9612232}{{\ttfamily arXiv:hep-ph/9612232}}.

\bibitem{Barr:1997hq}
S.~M. Barr and S.~Raby, ``{Minimal SO(10) unification},''
  \href{http://dx.doi.org/10.1103/PhysRevLett.79.4748}{{\em Phys. Rev. Lett.}
  {\bfseries 79} (1997) 4748--4751},
  \href{http://arxiv.org/abs/hep-ph/9705366}{{\ttfamily arXiv:hep-ph/9705366}}.

\bibitem{Chacko:1998jz}
Z.~Chacko and R.~N. Mohapatra, ``{Economical doublet triplet splitting and
  strong suppression of proton decay in SO(10)},''
  \href{http://dx.doi.org/10.1103/PhysRevD.59.011702}{{\em Phys. Rev. D}
  {\bfseries 59} (1999) 011702},
  \href{http://arxiv.org/abs/hep-ph/9808458}{{\ttfamily arXiv:hep-ph/9808458}}.

\bibitem{Babu:1998wi}
K.~S. Babu, J.~C. Pati, and F.~Wilczek, ``{Fermion masses, neutrino
  oscillations, and proton decay in the light of Super-Kamiokande},''
  \href{http://dx.doi.org/10.1016/S0550-3213(99)00589-1}{{\em Nucl. Phys. B}
  {\bfseries 566} (2000) 33--91},
  \href{http://arxiv.org/abs/hep-ph/9812538}{{\ttfamily arXiv:hep-ph/9812538}}.

\bibitem{Babu:2002fsa}
K.~S. Babu and S.~M. Barr, ``{Eliminating the d = 5 proton decay operators from
  SUSY GUTs},'' \href{http://dx.doi.org/10.1103/PhysRevD.65.095009}{{\em Phys.
  Rev. D} {\bfseries 65} (2002) 095009},
  \href{http://arxiv.org/abs/hep-ph/0201130}{{\ttfamily arXiv:hep-ph/0201130}}.

\bibitem{Kyae:2005vg}
B.~Kyae and Q.~Shafi, ``{Inflation with realistic supersymmetric SO(10)},''
  \href{http://dx.doi.org/10.1103/PhysRevD.72.063515}{{\em Phys. Rev. D}
  {\bfseries 72} (2005) 063515},
  \href{http://arxiv.org/abs/hep-ph/0504044}{{\ttfamily arXiv:hep-ph/0504044}}.

\bibitem{Babu:2010ej}
K.~S. Babu, J.~C. Pati, and Z.~Tavartkiladze, ``{Constraining Proton Lifetime
  in SO(10) with Stabilized Doublet-Triplet Splitting},''
  \href{http://dx.doi.org/10.1007/JHEP06(2010)084}{{\em JHEP} {\bfseries 06}
  (2010) 084}, \href{http://arxiv.org/abs/1003.2625}{{\ttfamily arXiv:1003.2625
  [hep-ph]}}.

\bibitem{Wan:2022glq}
Q.~Wan and D.-X. Zhang, ``{An extended study on the supersymmetric SO(10)
  models with natural doublet-triplet splitting},''
  \href{http://arxiv.org/abs/2210.01589}{{\ttfamily arXiv:2210.01589
  [hep-ph]}}.

\bibitem{Linde:1993cn}
A.~D. Linde, ``{Hybrid inflation},''
  \href{http://dx.doi.org/10.1103/PhysRevD.49.748}{{\em Phys. Rev. D}
  {\bfseries 49} (1994) 748--754},
  \href{http://arxiv.org/abs/astro-ph/9307002}{{\ttfamily
  arXiv:astro-ph/9307002}}.

\bibitem{linde1991axions}
A.~Linde, ``Axions in inflationary cosmology,'' {\em Physics Letters B}
  {\bfseries 259} no.~1-2, (1991) 38--47.

\bibitem{Dvali:1994ms}
G.~R. Dvali, Q.~Shafi, and R.~K. Schaefer, ``{Large scale structure and
  supersymmetric inflation without fine tuning},''
  \href{http://dx.doi.org/10.1103/PhysRevLett.73.1886}{{\em Phys. Rev. Lett.}
  {\bfseries 73} (1994) 1886--1889},
  \href{http://arxiv.org/abs/hep-ph/9406319}{{\ttfamily arXiv:hep-ph/9406319}}.

\bibitem{Antusch:2004hd}
S.~Antusch, M.~Bastero-Gil, S.~F. King, and Q.~Shafi, ``{Sneutrino hybrid
  inflation in supergravity},''
  \href{http://dx.doi.org/10.1103/PhysRevD.71.083519}{{\em Phys. Rev. D}
  {\bfseries 71} (2005) 083519},
  \href{http://arxiv.org/abs/hep-ph/0411298}{{\ttfamily arXiv:hep-ph/0411298}}.

\bibitem{Antusch:2010va}
S.~Antusch, M.~Bastero-Gil, J.~P. Baumann, K.~Dutta, S.~F. King, and P.~M.
  Kostka, ``{Gauge Non-Singlet Inflation in SUSY GUTs},''
  \href{http://dx.doi.org/10.1007/JHEP08(2010)100}{{\em JHEP} {\bfseries 08}
  (2010) 100}, \href{http://arxiv.org/abs/1003.3233}{{\ttfamily arXiv:1003.3233
  [hep-ph]}}.

\bibitem{Note1}
As a result, the appearance of automatic R-parity from within the SO(10) group
  is no longer possible. However a discrete symmetry, such as a $Z_2$ symmetry
  (matter parity), can readily be imposed.

\bibitem{Vachaspati:1984gt}
T.~Vachaspati and A.~Vilenkin, ``{Gravitational Radiation from Cosmic
  Strings},'' \href{http://dx.doi.org/10.1103/PhysRevD.31.3052}{{\em Phys. Rev.
  D} {\bfseries 31} (1985) 3052}.

\bibitem{Langacker:1980kd}
P.~Langacker and S.-Y. Pi, ``{Magnetic Monopoles in Grand Unified Theories},''
  \href{http://dx.doi.org/10.1103/PhysRevLett.45.1}{{\em Phys. Rev. Lett.}
  {\bfseries 45} (1980) 1}.

\bibitem{Lazarides:1981fv}
G.~Lazarides, Q.~Shafi, and T.~F. Walsh, ``{Cosmic Strings and Domains in
  Unified Theories},''
  \href{http://dx.doi.org/10.1016/0550-3213(82)90052-9}{{\em Nucl. Phys. B}
  {\bfseries 195} (1982) 157--172}.

\bibitem{Vilenkin:1982hm}
A.~Vilenkin, ``{Cosmological Evolution og Monopoles Connected by Strings},''
  \href{http://dx.doi.org/10.1016/0550-3213(82)90037-2}{{\em Nucl. Phys. B}
  {\bfseries 196} (1982) 240--258}.

\bibitem{Leblond:2009fq}
L.~Leblond, B.~Shlaer, and X.~Siemens, ``{Gravitational Waves from Broken
  Cosmic Strings: The Bursts and the Beads},''
  \href{http://dx.doi.org/10.1103/PhysRevD.79.123519}{{\em Phys. Rev. D}
  {\bfseries 79} (2009) 123519},
  \href{http://arxiv.org/abs/0903.4686}{{\ttfamily arXiv:0903.4686
  [astro-ph.CO]}}.

\bibitem{Auclair:2019wcv}
P.~Auclair {\em et~al.}, ``{Probing the gravitational wave background from
  cosmic strings with LISA},''
  \href{http://dx.doi.org/10.1088/1475-7516/2020/04/034}{{\em JCAP} {\bfseries
  04} (2020) 034}, \href{http://arxiv.org/abs/1909.00819}{{\ttfamily
  arXiv:1909.00819 [astro-ph.CO]}}.

\bibitem{Cui:2018rwi}
Y.~Cui, M.~Lewicki, D.~E. Morrissey, and J.~D. Wells, ``{Probing the pre-BBN
  universe with gravitational waves from cosmic strings},''
  \href{http://dx.doi.org/10.1007/JHEP01(2019)081}{{\em JHEP} {\bfseries 01}
  (2019) 081}, \href{http://arxiv.org/abs/1808.08968}{{\ttfamily
  arXiv:1808.08968 [hep-ph]}}.

\bibitem{Blasi:2020wpy}
S.~Blasi, V.~Brdar, and K.~Schmitz, ``{Fingerprint of low-scale leptogenesis in
  the primordial gravitational-wave spectrum},''
  \href{http://dx.doi.org/10.1103/PhysRevResearch.2.043321}{{\em Phys. Rev.
  Res.} {\bfseries 2} no.~4, (2020) 043321},
  \href{http://arxiv.org/abs/2004.02889}{{\ttfamily arXiv:2004.02889
  [hep-ph]}}.

\bibitem{Note2}
Stable cosmic strings, however, were consistent with the previous PTA data. For
  works on GWs, in light of NANOGrav12.5 data, arising from cosmic strings
  within GUTs, c.f.,~\cite
  {Buchmuller:2019gfy,King:2020hyd,King:2021gmj,Lazarides:2022jgr,Fu:2022lrn,Saad:2022mzu,Lazarides:2023iim,Maji:2023fba,Madge:2023cak}.

\bibitem{Buchmuller:2021mbb}
W.~Buchmuller, V.~Domcke, and K.~Schmitz, ``{Stochastic gravitational-wave
  background from metastable cosmic strings},''
  \href{http://dx.doi.org/10.1088/1475-7516/2021/12/006}{{\em JCAP} {\bfseries
  12} no.~12, (2021) 006}, \href{http://arxiv.org/abs/2107.04578}{{\ttfamily
  arXiv:2107.04578 [hep-ph]}}.

\bibitem{LIGOScientific:2021nrg}
{\bfseries LIGO Scientific, Virgo, KAGRA} Collaboration, R.~Abbott {\em
  et~al.}, ``{Constraints on Cosmic Strings Using Data from the Third Advanced
  LIGO\textendash{}Virgo Observing Run},''
  \href{http://dx.doi.org/10.1103/PhysRevLett.126.241102}{{\em Phys. Rev.
  Lett.} {\bfseries 126} no.~24, (2021) 241102},
  \href{http://arxiv.org/abs/2101.12248}{{\ttfamily arXiv:2101.12248 [gr-qc]}}.

\bibitem{KAGRA:2021kbb}
{\bfseries KAGRA, Virgo, LIGO Scientific} Collaboration, R.~Abbott {\em
  et~al.}, ``{Upper limits on the isotropic gravitational-wave background from
  Advanced LIGO and Advanced Virgo\textquoteright{}s third observing run},''
  \href{http://dx.doi.org/10.1103/PhysRevD.104.022004}{{\em Phys. Rev. D}
  {\bfseries 104} no.~2, (2021) 022004},
  \href{http://arxiv.org/abs/2101.12130}{{\ttfamily arXiv:2101.12130 [gr-qc]}}.

\bibitem{Janssen:2014dka}
G.~Janssen {\em et~al.}, ``{Gravitational wave astronomy with the SKA},''
  \href{http://dx.doi.org/10.22323/1.215.0037}{{\em PoS} {\bfseries AASKA14}
  (2015) 037}, \href{http://arxiv.org/abs/1501.00127}{{\ttfamily
  arXiv:1501.00127 [astro-ph.IM]}}.

\bibitem{Theia:2017xtk}
{\bfseries Theia} Collaboration, C.~Boehm {\em et~al.}, ``{Theia: Faint objects
  in motion or the new astrometry frontier},''
  \href{http://arxiv.org/abs/1707.01348}{{\ttfamily arXiv:1707.01348
  [astro-ph.IM]}}.

\bibitem{Sesana:2019vho}
A.~Sesana {\em et~al.}, ``{Unveiling the gravitational universe at $\mu$-Hz
  frequencies},'' \href{http://dx.doi.org/10.1007/s10686-021-09709-9}{{\em
  Exper. Astron.} {\bfseries 51} no.~3, (2021) 1333--1383},
  \href{http://arxiv.org/abs/1908.11391}{{\ttfamily arXiv:1908.11391
  [astro-ph.IM]}}.

\bibitem{Audley:2017drz}
{\bfseries LISA} Collaboration, P.~Amaro-Seoane {\em et~al.}, ``{Laser
  Interferometer Space Antenna},''
  \href{http://arxiv.org/abs/1702.00786}{{\ttfamily arXiv:1702.00786
  [astro-ph.IM]}}.

\bibitem{Guo:2018npi}
W.-H. Ruan, Z.-K. Guo, R.-G. Cai, and Y.-Z. Zhang, ``{Taiji program:
  Gravitational-wave sources},''
  \href{http://dx.doi.org/10.1142/S0217751X2050075X}{{\em Int. J. Mod. Phys. A}
  {\bfseries 35} no.~17, (2020) 2050075},
  \href{http://arxiv.org/abs/1807.09495}{{\ttfamily arXiv:1807.09495 [gr-qc]}}.

\bibitem{Luo:2015ght}
{\bfseries TianQin} Collaboration, J.~Luo {\em et~al.}, ``{TianQin: a
  space-borne gravitational wave detector},''
  \href{http://dx.doi.org/10.1088/0264-9381/33/3/035010}{{\em Class. Quant.
  Grav.} {\bfseries 33} no.~3, (2016) 035010},
  \href{http://arxiv.org/abs/1512.02076}{{\ttfamily arXiv:1512.02076
  [astro-ph.IM]}}.

\bibitem{Corbin:2005ny}
V.~Corbin and N.~J. Cornish, ``{Detecting the cosmic gravitational wave
  background with the big bang observer},''
  \href{http://dx.doi.org/10.1088/0264-9381/23/7/014}{{\em Class. Quant. Grav.}
  {\bfseries 23} (2006) 2435--2446},
  \href{http://arxiv.org/abs/gr-qc/0512039}{{\ttfamily arXiv:gr-qc/0512039}}.

\bibitem{Seto:2001qf}
N.~Seto, S.~Kawamura, and T.~Nakamura, ``{Possibility of direct measurement of
  the acceleration of the universe using 0.1-Hz band laser interferometer
  gravitational wave antenna in space},''
  \href{http://dx.doi.org/10.1103/PhysRevLett.87.221103}{{\em Phys. Rev. Lett.}
  {\bfseries 87} (2001) 221103},
  \href{http://arxiv.org/abs/astro-ph/0108011}{{\ttfamily
  arXiv:astro-ph/0108011}}.

\bibitem{Sathyaprakash:2012jk}
B.~Sathyaprakash {\em et~al.}, ``{Scientific Objectives of Einstein
  Telescope},'' \href{http://dx.doi.org/10.1088/0264-9381/29/12/124013}{{\em
  Class. Quant. Grav.} {\bfseries 29} (2012) 124013},
  \href{http://arxiv.org/abs/1206.0331}{{\ttfamily arXiv:1206.0331 [gr-qc]}}.
  [Erratum: Class.Quant.Grav. 30, 079501 (2013)].

\bibitem{Evans:2016mbw}
{\bfseries LIGO Scientific} Collaboration, B.~P. Abbott {\em et~al.},
  ``{Exploring the Sensitivity of Next Generation Gravitational Wave
  Detectors},'' \href{http://dx.doi.org/10.1088/1361-6382/aa51f4}{{\em Class.
  Quant. Grav.} {\bfseries 34} no.~4, (2017) 044001},
  \href{http://arxiv.org/abs/1607.08697}{{\ttfamily arXiv:1607.08697
  [astro-ph.IM]}}.

\bibitem{James:1992zp}
M.~James, L.~Perivolaropoulos, and T.~Vachaspati, ``{Stability of electroweak
  strings},'' \href{http://dx.doi.org/10.1103/PhysRevD.46.R5232}{{\em Phys.
  Rev. D} {\bfseries 46} (1992) R5232--R5235}.

\bibitem{James:1992wb}
M.~James, L.~Perivolaropoulos, and T.~Vachaspati, ``{Detailed stability
  analysis of electroweak strings},''
  \href{http://dx.doi.org/10.1016/0550-3213(93)90046-R}{{\em Nucl. Phys. B}
  {\bfseries 395} (1993) 534--546},
  \href{http://arxiv.org/abs/hep-ph/9212301}{{\ttfamily arXiv:hep-ph/9212301}}.

\bibitem{Goodband:1995he}
M.~Goodband and M.~Hindmarsh, ``{Instabilities of electroweak strings},''
  \href{http://dx.doi.org/10.1016/0370-2693(95)01198-Y}{{\em Phys. Lett. B}
  {\bfseries 363} (1995) 58--64},
  \href{http://arxiv.org/abs/hep-ph/9505357}{{\ttfamily arXiv:hep-ph/9505357}}.

\bibitem{Jeannerot:2003qv}
R.~Jeannerot, J.~Rocher, and M.~Sakellariadou, ``{How generic is cosmic string
  formation in SUSY GUTs},''
  \href{http://dx.doi.org/10.1103/PhysRevD.68.103514}{{\em Phys. Rev. D}
  {\bfseries 68} (2003) 103514},
  \href{http://arxiv.org/abs/hep-ph/0308134}{{\ttfamily arXiv:hep-ph/0308134}}.

\bibitem{Zu:2023olm}
L.~Zu, C.~Zhang, Y.-Y. Li, Y.-C. Gu, Y.-L.~S. Tsai, and Y.-Z. Fan, ``{Mirror
  QCD phase transition as the origin of the nanohertz Stochastic
  Gravitational-Wave Background detected by the Pulsar Timing Arrays},''
  \href{http://arxiv.org/abs/2306.16769}{{\ttfamily arXiv:2306.16769
  [astro-ph.HE]}}.

\bibitem{Vagnozzi:2023lwo}
S.~Vagnozzi, ``{Inflationary interpretation of the stochastic gravitational
  wave background signal detected by pulsar timing array experiments},''
  \href{http://arxiv.org/abs/2306.16912}{{\ttfamily arXiv:2306.16912
  [astro-ph.CO]}}.

\bibitem{Han:2023olf}
C.~Han, K.-P. Xie, J.~M. Yang, and M.~Zhang, ``{Self-interacting dark matter
  implied by nano-Hertz gravitational waves},''
  \href{http://arxiv.org/abs/2306.16966}{{\ttfamily arXiv:2306.16966
  [hep-ph]}}.

\bibitem{Lambiase:2023pxd}
G.~Lambiase, L.~Mastrototaro, and L.~Visinelli, ``{Astrophysical neutrino
  oscillations after pulsar timing array analyses},''
  \href{http://arxiv.org/abs/2306.16977}{{\ttfamily arXiv:2306.16977
  [astro-ph.HE]}}.

\bibitem{Ellis:2023dgf}
J.~Ellis, M.~Fairbairn, G.~H\"utsi, J.~Raidal, J.~Urrutia, V.~Vaskonen, and
  H.~Veerm\"ae, ``{Gravitational Waves from SMBH Binaries in Light of the
  NANOGrav 15-Year Data},'' \href{http://arxiv.org/abs/2306.17021}{{\ttfamily
  arXiv:2306.17021 [astro-ph.CO]}}.

\bibitem{Guo:2023hyp}
S.-Y. Guo, M.~Khlopov, X.~Liu, L.~Wu, Y.~Wu, and B.~Zhu, ``{Footprints of
  Axion-Like Particle in Pulsar Timing Array Data and JWST Observations},''
  \href{http://arxiv.org/abs/2306.17022}{{\ttfamily arXiv:2306.17022
  [hep-ph]}}.

\bibitem{Megias:2023kiy}
E.~Megias, G.~Nardini, and M.~Quiros, ``{Pulsar Timing Array Stochastic
  Background from light Kaluza-Klein resonances},''
  \href{http://arxiv.org/abs/2306.17071}{{\ttfamily arXiv:2306.17071
  [hep-ph]}}.

\bibitem{Fujikura:2023lkn}
K.~Fujikura, S.~Girmohanta, Y.~Nakai, and M.~Suzuki, ``{NANOGrav Signal from a
  Dark Conformal Phase Transition},''
  \href{http://arxiv.org/abs/2306.17086}{{\ttfamily arXiv:2306.17086
  [hep-ph]}}.

\bibitem{Yang:2023aak}
J.~Yang, N.~Xie, and F.~P. Huang, ``{Nano-Hertz stochastic gravitational wave
  background as hints of ultralight axion particles},''
  \href{http://arxiv.org/abs/2306.17113}{{\ttfamily arXiv:2306.17113
  [hep-ph]}}.

\bibitem{Franciolini:2023wjm}
G.~Franciolini, D.~Racco, and F.~Rompineve, ``{Footprints of the QCD Crossover
  on Cosmological Gravitational Waves at Pulsar Timing Arrays},''
  \href{http://arxiv.org/abs/2306.17136}{{\ttfamily arXiv:2306.17136
  [astro-ph.CO]}}.

\bibitem{Shen:2023pan}
Z.-Q. Shen, G.-W. Yuan, Y.-Y. Wang, and Y.-Z. Wang, ``{Dark Matter Spike
  surrounding Supermassive Black Holes Binary and the nanohertz Stochastic
  Gravitational Wave Background},''
  \href{http://arxiv.org/abs/2306.17143}{{\ttfamily arXiv:2306.17143
  [astro-ph.HE]}}.

\bibitem{Kitajima:2023cek}
N.~Kitajima, J.~Lee, K.~Murai, F.~Takahashi, and W.~Yin, ``{Nanohertz
  Gravitational Waves from Axion Domain Walls Coupled to QCD},''
  \href{http://arxiv.org/abs/2306.17146}{{\ttfamily arXiv:2306.17146
  [hep-ph]}}.

\bibitem{Ellis:2023tsl}
J.~Ellis, M.~Lewicki, C.~Lin, and V.~Vaskonen, ``{Cosmic Superstrings Revisited
  in Light of NANOGrav 15-Year Data},''
  \href{http://arxiv.org/abs/2306.17147}{{\ttfamily arXiv:2306.17147
  [astro-ph.CO]}}.

\bibitem{Franciolini:2023pbf}
G.~Franciolini, A.~Iovino, Junior., V.~Vaskonen, and H.~Veermae, ``{The recent
  gravitational wave observation by pulsar timing arrays and primordial black
  holes: the importance of non-gaussianities},''
  \href{http://arxiv.org/abs/2306.17149}{{\ttfamily arXiv:2306.17149
  [astro-ph.CO]}}.

\bibitem{Ghoshal:2023fhh}
A.~Ghoshal and A.~Strumia, ``{Probing the Dark Matter density with
  gravitational waves from super-massive binary black holes},''
  \href{http://arxiv.org/abs/2306.17158}{{\ttfamily arXiv:2306.17158
  [astro-ph.CO]}}.

\bibitem{Bai:2023cqj}
Y.~Bai, T.-K. Chen, and M.~Korwar, ``{QCD-Collapsed Domain Walls: QCD Phase
  Transition and Gravitational Wave Spectroscopy},''
  \href{http://arxiv.org/abs/2306.17160}{{\ttfamily arXiv:2306.17160
  [hep-ph]}}.

\bibitem{Addazi:2023jvg}
A.~Addazi, Y.-F. Cai, A.~Marciano, and L.~Visinelli, ``{Have pulsar timing
  array methods detected a cosmological phase transition?},''
  \href{http://arxiv.org/abs/2306.17205}{{\ttfamily arXiv:2306.17205
  [astro-ph.CO]}}.

\bibitem{Athron:2023mer}
P.~Athron, A.~Fowlie, C.-T. Lu, L.~Morris, L.~Wu, Y.~Wu, and Z.~Xu, ``{Can
  Supercooled Phase Transitions explain the Gravitational Wave Background
  observed by Pulsar Timing Arrays?},''
  \href{http://arxiv.org/abs/2306.17239}{{\ttfamily arXiv:2306.17239
  [hep-ph]}}.

\bibitem{Kitajima:2023vre}
N.~Kitajima and K.~Nakayama, ``{Nanohertz gravitational waves from cosmic
  strings and dark photon dark matter},''
  \href{http://arxiv.org/abs/2306.17390}{{\ttfamily arXiv:2306.17390
  [hep-ph]}}.

\bibitem{Lazarides:2023ksx}
G.~Lazarides, R.~Maji, and Q.~Shafi, ``{Superheavy quasi-stable strings and
  walls bounded by strings in the light of NANOGrav 15 year data},''
  \href{http://arxiv.org/abs/2306.17788}{{\ttfamily arXiv:2306.17788
  [hep-ph]}}.

\bibitem{Broadhurst:2023tus}
T.~Broadhurst, C.~Chen, T.~Liu, and K.-F. Zheng, ``{Binary Supermassive Black
  Holes Orbiting Dark Matter Solitons: From the Dual AGN in UGC4211 to
  NanoHertz Gravitational Waves},''
  \href{http://arxiv.org/abs/2306.17821}{{\ttfamily arXiv:2306.17821
  [astro-ph.HE]}}.

\bibitem{Cai:2023dls}
Y.-F. Cai, X.-C. He, X.~Ma, S.-F. Yan, and G.-W. Yuan, ``{Limits on
  scalar-induced gravitational waves from the stochastic background by pulsar
  timing array observations},''
  \href{http://arxiv.org/abs/2306.17822}{{\ttfamily arXiv:2306.17822 [gr-qc]}}.

\bibitem{Yang:2023qlf}
A.~Yang, J.~Ma, S.~Jiang, and F.~P. Huang, ``{Implication of nano-Hertz
  stochastic gravitational wave on dynamical dark matter through a first-order
  phase transition},'' \href{http://arxiv.org/abs/2306.17827}{{\ttfamily
  arXiv:2306.17827 [hep-ph]}}.

\bibitem{Blasi:2023sej}
S.~Blasi, A.~Mariotti, A.~Rase, and A.~Sevrin, ``{Axionic domain walls at
  Pulsar Timing Arrays: QCD bias and particle friction},''
  \href{http://arxiv.org/abs/2306.17830}{{\ttfamily arXiv:2306.17830
  [hep-ph]}}.

\bibitem{Inomata:2023zup}
K.~Inomata, K.~Kohri, and T.~Terada, ``{The Detected Stochastic Gravitational
  Waves and Sub-Solar Primordial Black Holes},''
  \href{http://arxiv.org/abs/2306.17834}{{\ttfamily arXiv:2306.17834
  [astro-ph.CO]}}.

\bibitem{Depta:2023qst}
P.~F. Depta, K.~Schmidt-Hoberg, and C.~Tasillo, ``{Do pulsar timing arrays
  observe merging primordial black holes?},''
  \href{http://arxiv.org/abs/2306.17836}{{\ttfamily arXiv:2306.17836
  [astro-ph.CO]}}.

\bibitem{Gouttenoire:2023ftk}
Y.~Gouttenoire and E.~Vitagliano, ``{Domain wall interpretation of the PTA
  signal confronting black hole overproduction},''
  \href{http://arxiv.org/abs/2306.17841}{{\ttfamily arXiv:2306.17841 [gr-qc]}}.

\bibitem{Borah:2023sbc}
D.~Borah, S.~Jyoti~Das, and R.~Samanta, ``{Inflationary origin of gravitational
  waves with \textbackslash{}textit{Miracle-less WIMP} dark matter in the light
  of recent PTA results},'' \href{http://arxiv.org/abs/2307.00537}{{\ttfamily
  arXiv:2307.00537 [hep-ph]}}.

\bibitem{Murai:2023gkv}
K.~Murai and W.~Yin, ``{A Novel Probe of Supersymmetry in Light of Nanohertz
  Gravitational Waves},'' \href{http://arxiv.org/abs/2307.00628}{{\ttfamily
  arXiv:2307.00628 [hep-ph]}}.

\bibitem{Datta:2023vbs}
S.~Datta, ``{Inflationary gravitational waves, pulsar timing data and
  low-scale-leptogenesis},'' \href{http://arxiv.org/abs/2307.00646}{{\ttfamily
  arXiv:2307.00646 [hep-ph]}}.

\bibitem{Barman:2023fad}
B.~Barman, D.~Borah, S.~Jyoti~Das, and I.~Saha, ``{Scale of Dirac leptogenesis
  and left-right symmetry in the light of recent PTA results},''
  \href{http://arxiv.org/abs/2307.00656}{{\ttfamily arXiv:2307.00656
  [hep-ph]}}.

\bibitem{Bi:2023tib}
Y.-C. Bi, Y.-M. Wu, Z.-C. Chen, and Q.-G. Huang, ``{Implications for the
  Supermassive Black Hole Binaries from the NANOGrav 15-year Data Set},''
  \href{http://arxiv.org/abs/2307.00722}{{\ttfamily arXiv:2307.00722
  [astro-ph.CO]}}.

\bibitem{Lu:2023mcz}
B.-Q. Lu and C.-W. Chiang, ``{Nano-Hertz stochastic gravitational wave
  background from domain wall annihilation},''
  \href{http://arxiv.org/abs/2307.00746}{{\ttfamily arXiv:2307.00746
  [hep-ph]}}.

\bibitem{Xiao:2023dbb}
Y.~Xiao, J.~M. Yang, and Y.~Zhang, ``{Implications of Nano-Hertz Gravitational
  Waves on Electroweak Phase Transition in the Singlet Dark Matter Model},''
  \href{http://arxiv.org/abs/2307.01072}{{\ttfamily arXiv:2307.01072
  [hep-ph]}}.

\bibitem{Li:2023bxy}
S.-P. Li and K.-P. Xie, ``{A collider test of nano-Hertz gravitational waves
  from pulsar timing arrays},''
  \href{http://arxiv.org/abs/2307.01086}{{\ttfamily arXiv:2307.01086
  [hep-ph]}}.

\bibitem{Anchordoqui:2023tln}
L.~A. Anchordoqui, I.~Antoniadis, and D.~Lust, ``{Fuzzy Dark Matter, the Dark
  Dimension, and the Pulsar Timing Array Signal},''
  \href{http://arxiv.org/abs/2307.01100}{{\ttfamily arXiv:2307.01100
  [hep-ph]}}.

\bibitem{Niu:2023bsr}
X.~Niu and M.~H. Rahat, ``{NANOGrav signal from axion inflation},''
  \href{http://arxiv.org/abs/2307.01192}{{\ttfamily arXiv:2307.01192
  [hep-ph]}}.

\bibitem{Ebadi:2023xhq}
R.~Ebadi, S.~Kumar, A.~McCune, H.~Tai, and L.-T. Wang, ``{Gravitational Waves
  from Stochastic Scalar Fluctuations},''
  \href{http://arxiv.org/abs/2307.01248}{{\ttfamily arXiv:2307.01248
  [astro-ph.CO]}}.

\bibitem{Gouttenoire:2023nzr}
Y.~Gouttenoire, S.~Trifinopoulos, G.~Valogiannis, and M.~Vanvlasselaer,
  ``{Scrutinizing the Primordial Black Holes Interpretation of PTA
  Gravitational Waves and JWST Early Galaxies},''
  \href{http://arxiv.org/abs/2307.01457}{{\ttfamily arXiv:2307.01457
  [astro-ph.CO]}}.

\bibitem{Abe:2023yrw}
K.~T. Abe and Y.~Tada, ``{Translating nano-Hertz gravitational wave background
  into primordial perturbations taking account of the cosmological QCD phase
  transition},'' \href{http://arxiv.org/abs/2307.01653}{{\ttfamily
  arXiv:2307.01653 [astro-ph.CO]}}.

\bibitem{Ghosh:2023aum}
T.~Ghosh, A.~Ghoshal, H.-K. Guo, F.~Hajkarim, S.~F. King, K.~Sinha, X.~Wang,
  and G.~White, ``{Did we hear the sound of the Universe boiling? Analysis
  using the full fluid velocity profiles and NANOGrav 15-year data},''
  \href{http://arxiv.org/abs/2307.02259}{{\ttfamily arXiv:2307.02259
  [astro-ph.HE]}}.

\bibitem{Unal:2023srk}
C.~Unal, A.~Papageorgiou, and I.~Obata, ``{Axion-Gauge Dynamics During
  Inflation as the Origin of Pulsar Timing Array Signals and Primordial Black
  Holes},'' \href{http://arxiv.org/abs/2307.02322}{{\ttfamily arXiv:2307.02322
  [astro-ph.CO]}}.

\bibitem{Bian:2023dnv}
L.~Bian, S.~Ge, J.~Shu, B.~Wang, X.-Y. Yang, and J.~Zong, ``{Gravitational wave
  sources for Pulsar Timing Arrays},''
  \href{http://arxiv.org/abs/2307.02376}{{\ttfamily arXiv:2307.02376
  [astro-ph.HE]}}.

\bibitem{Figueroa:2023zhu}
D.~G. Figueroa, M.~Pieroni, A.~Ricciardone, and P.~Simakachorn, ``{Cosmological
  Background Interpretation of Pulsar Timing Array Data},''
  \href{http://arxiv.org/abs/2307.02399}{{\ttfamily arXiv:2307.02399
  [astro-ph.CO]}}.

\bibitem{Du:2023qvj}
X.~K. Du, M.~X. Huang, F.~Wang, and Y.~K. Zhang, ``{Did the nHZ Gravitational
  Waves Signatures Observed By NANOGrav Indicate Multiple Sector SUSY
  Breaking?},'' \href{http://arxiv.org/abs/2307.02938}{{\ttfamily
  arXiv:2307.02938 [hep-ph]}}.

\bibitem{Servant:2023mwt}
G.~Servant and P.~Simakachorn, ``{Constraining Post-Inflationary Axions with
  Pulsar Timing Arrays},'' \href{http://arxiv.org/abs/2307.03121}{{\ttfamily
  arXiv:2307.03121 [hep-ph]}}.

\bibitem{Li:2023tdx}
X.-F. Li, ``{Probing the high temperature symmetry breaking with gravitational
  waves from domain walls},'' \href{http://arxiv.org/abs/2307.03163}{{\ttfamily
  arXiv:2307.03163 [hep-ph]}}.

\bibitem{DiBari:2023upq}
P.~Di~Bari and M.~H. Rahat, ``{The split majoron model confronts the NANOGrav
  signal},'' \href{http://arxiv.org/abs/2307.03184}{{\ttfamily arXiv:2307.03184
  [hep-ph]}}.

\bibitem{Geller:2023shn}
M.~Geller, S.~Ghosh, S.~Lu, and Y.~Tsai, ``{Challenges in Interpreting the
  NANOGrav 15-Year Data Set as Early Universe Gravitational Waves Produced by
  ALP Induced Instability},'' \href{http://arxiv.org/abs/2307.03724}{{\ttfamily
  arXiv:2307.03724 [hep-ph]}}.

\bibitem{Liu:2023ymk}
L.~Liu, Z.-C. Chen, and Q.-G. Huang, ``{Implications for the non-Gaussianity of
  curvature perturbation from pulsar timing arrays},''
  \href{http://arxiv.org/abs/2307.01102}{{\ttfamily arXiv:2307.01102
  [astro-ph.CO]}}.

\bibitem{Buchmuller:2019gfy}
W.~Buchm\"uller, V.~Domcke, H.~Murayama, and K.~Schmitz, ``{Probing the scale
  of grand unification with gravitational waves},''
  \href{http://dx.doi.org/10.1016/j.physletb.2020.135764}{{\em Phys. Lett. B}
  {\bfseries 809} (2020) 135764},
  \href{http://arxiv.org/abs/1912.03695}{{\ttfamily arXiv:1912.03695
  [hep-ph]}}.

\bibitem{King:2020hyd}
S.~F. King, S.~Pascoli, J.~Turner, and Y.-L. Zhou, ``{Gravitational Waves and
  Proton Decay: Complementary Windows into Grand Unified Theories},''
  \href{http://dx.doi.org/10.1103/PhysRevLett.126.021802}{{\em Phys. Rev.
  Lett.} {\bfseries 126} no.~2, (2021) 021802},
  \href{http://arxiv.org/abs/2005.13549}{{\ttfamily arXiv:2005.13549
  [hep-ph]}}.

\bibitem{King:2021gmj}
S.~F. King, S.~Pascoli, J.~Turner, and Y.-L. Zhou, ``{Confronting SO(10) GUTs
  with proton decay and gravitational waves},''
  \href{http://dx.doi.org/10.1007/JHEP10(2021)225}{{\em JHEP} {\bfseries 10}
  (2021) 225}, \href{http://arxiv.org/abs/2106.15634}{{\ttfamily
  arXiv:2106.15634 [hep-ph]}}.

\bibitem{Lazarides:2022jgr}
G.~Lazarides, R.~Maji, and Q.~Shafi, ``{Gravitational waves from quasi-stable
  strings},'' \href{http://dx.doi.org/10.1088/1475-7516/2022/08/042}{{\em JCAP}
  {\bfseries 08} no.~08, (2022) 042},
  \href{http://arxiv.org/abs/2203.11204}{{\ttfamily arXiv:2203.11204
  [hep-ph]}}.

\bibitem{Fu:2022lrn}
B.~Fu, S.~F. King, L.~Marsili, S.~Pascoli, J.~Turner, and Y.-L. Zhou, ``{A
  predictive and testable unified theory of fermion masses, mixing and
  leptogenesis},'' \href{http://dx.doi.org/10.1007/JHEP11(2022)072}{{\em JHEP}
  {\bfseries 11} (2022) 072}, \href{http://arxiv.org/abs/2209.00021}{{\ttfamily
  arXiv:2209.00021 [hep-ph]}}.

\bibitem{Saad:2022mzu}
S.~Saad, ``{Probing minimal grand unification through gravitational waves,
  proton decay, and fermion masses},''
  \href{http://dx.doi.org/10.1007/JHEP04(2023)058}{{\em JHEP} {\bfseries 04}
  (2023) 058}, \href{http://arxiv.org/abs/2212.05291}{{\ttfamily
  arXiv:2212.05291 [hep-ph]}}.

\bibitem{Lazarides:2023iim}
G.~Lazarides, Q.~Shafi, and A.~Tiwari, ``{Composite topological structures in
  SO(10)},'' \href{http://dx.doi.org/10.1007/JHEP05(2023)119}{{\em JHEP}
  {\bfseries 05} (2023) 119}, \href{http://arxiv.org/abs/2303.15159}{{\ttfamily
  arXiv:2303.15159 [hep-ph]}}.

\bibitem{Maji:2023fba}
R.~Maji, W.-I. Park, and Q.~Shafi, ``{Gravitational waves from walls bounded by
  strings in $SO(10)$ model of pseudo-Goldstone dark matter},''
  \href{http://arxiv.org/abs/2305.11775}{{\ttfamily arXiv:2305.11775
  [hep-ph]}}.

\bibitem{Madge:2023cak}
E.~Madge, E.~Morgante, C.~Puchades-Ib\'a\~nez, N.~Ramberg, W.~Ratzinger,
  S.~Schenk, and P.~Schwaller, ``{Primordial gravitational waves in the
  nano-Hertz regime and PTA data -- towards solving the GW inverse problem},''
  \href{http://arxiv.org/abs/2306.14856}{{\ttfamily arXiv:2306.14856
  [hep-ph]}}.

\end{thebibliography}\endgroup
\end{document}